# Characteristics of superconducting single photon detector in DPS-QKD system under bright illumination blinding attack

Mikio Fujiwara,[1,*] Toshimori Honjo,[2] Kaoru Shimizu,[3] Kiyoshi Tamaki,[3] and Masahide Sasaki[1]

[1] *National Institute of Information and Communication Technology, 4-2-1 Nukui-kitamachi, Koganei, Tokyo 184-8795, Japan*
[2] *NTT Secure Platform Laboratories, NTT Corporation, 3-9-11 Midori-cho, Musashino, Tokyo 180-8585, Japan*
[3] *NTT Basic Research Laboratories, NTT Corporation, 3-11 Morinosato Wakamiya, Atsugi, Kanagawa 180-8585, Japan*
[*] *fujiwara@nict.go.jp*

**Abstract:** We derive the time-dependent photo-detection probability equation of a superconducting single photon detector (SSPD) to study the responsive property for a pulse train at high repetition rate. Using this equation, we analyze the characteristics of SSPDs when illuminated by bright pulses in blinding attack on a quantum key distribution (QKD). We obtain good agreement between expected values based on our equation and actual experimental values. Such a time-dependent probability analysis contributes to security analysis.

**OCIS codes:** (270.5568) Quantum cryptography; (270.5570) Quantum detectors.

## 1. Introduction

Quantum key distribution (QKD) [1] allows two users, Alice and Bob, to share random numbers with the unconditional security guaranteed by the fundamental laws of physics. BB84[2] is the most famous protocol of QKD. Recently QKD systems have been deployed in the field environment, and tested for their reliability.[3-5] One should, however, note that unconditional security proofs assume that devices operate as required and their imperfections are within certain ranges as specified by the mathematical model. A gap between theory and practical implementation, i. e. side channels, can be reduced, but never vanishes. Protocols with simpler implementation are generally less vulnerable to side channels. Differential-phase-shift (DPS)-QKD protocol [6, 7] has a simpler structure, especially in the receiver, which consists of an asymmetric Mach-Zehnder interferometer and two single photon detectors as in Fig. 1(a), and hence enables stable operation and long distance transmission. It has been applied for 90 km transmission with SSPDs [8-11] in the Tokyo QKD network. [3] The SSPDs used there have very low dark count rates, and therefore they are suitable for a long distance QKD. Unfortunately, however, these detectors are threatened by detector side channel attacks.

Actually, eavesdropping by controlling single photon detectors (SPDs) have been proposed and demonstrated as mentioned later. In general, SPDs cannot count a photon number in the pulse, and they discriminate differential signals only, because SPDs are connected to amplifiers with capacitive coupling, and it follows that a user cannot know direct current level of SPDs. These facts imply that SPDs cannot distinguish between no-photon state and a state with no change which can be caused by continuously illuminating with bright input power beyond the normal operating range. Thus, it is possible to blind SPDs by bright continuous wave (CW) illumination. Moreover, semiconductor SPDs have "linear region" in which the output voltage from an SPD linearly increases with the power of very bright photo-illumination.[12, 13] Such a characteristics enables intercept and resend attack[14], without being detected by Alice and Bob. Recently, hacking of commercial QKD systems by CW illumination superposed with strong pulses have been reported.[13] On the other hand, SSPDs do not have "linear region" of output voltage to input photo-power. To disguise the single photon counting event of SSPDs, attackers must revive superconductivity in SSPDs after the bright illumination blinding. That means sequential pulse illumination with well-calibrated modulation (bright and dark) should be illuminated to SSPDs. Indeed, hacking SSPDs in DPS-QKD system by bright illumination with tailored power-reduced pulses was proposed[15] and demonstrated.[16]. In that case, the recovery time of SSPDs should be analyzed carefully. In general, the recovery times of SPDs in QKD systems are longer than pulse recurrence period of a QKD system. This is because it is difficult to develop a high rate SPD, and also the photo-detection rate is much smaller than photon transmission frequency due to the high attenuation in a transmission line and small photon number per pulse in QKD protocol. If the repetition period of the photo–detection is closer to the recovery time of a SPD, the detection efficiency changes according to history of photo-detection events. In addition, eavesdropping using pulse train named "sequential attack"[17] was proposed against the DPS-QKD system, where the detection events are bunched as a result of the eavesdropping. To catch this attack, the users must monitor the correlations of the detection events. Hence it is necessary to characterize the behavior of an SSPD against a pulse train to study loophole of the QKD system with SSPDs.

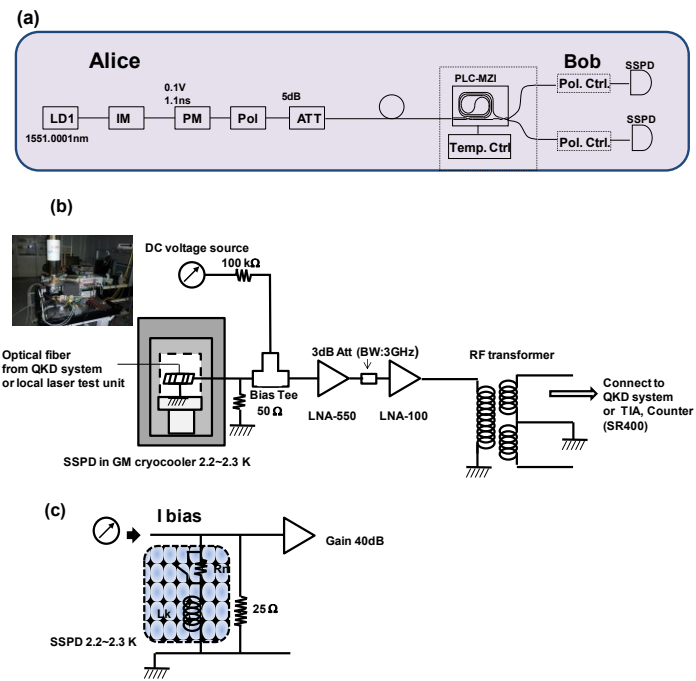

Fig. 1. (a) Conceptual view of DPS-QKD system. (b) Conceptual view and (c) equivalent circuit of the superconductor single photon detector (SSPD).

In the experiment in Ref. [15], CW illumination with power of a few mW was used to blind SSPDs. Note that blinding attack with such strong power illumination can be detected easily, for instance by tapping 1% of optical pulses from a quantum channel and measuring the power of them. In this power measurement, Bob can detect the blinding attack using a commercial optical-power meter, if CW illumination with the power of mW region was used in the hacking. On the other hand, Honjo et al.[16] demonstrated controlling SSPDs in the 1 GHz clocked DPS-QKD system and countermeasure against hacking by using pulse shaped bright illumination. They succeeded in decreasing the input optical-power to a few μW. However, it means that more than 10000 photons/pulse was needed to blind SSPDs. To know the behaviors of SSPDs in bright illumination is necessary to consider more effective way of bright illumination attack. In this paper, we discuss the equation of the photo-detection probability of our SSPDs to pulse train, and compare counting rate from the equation with experimental results. Such a consideration is useful for the effective hacking as well as countermeasure against bright illumination blinding attack on SSPDs.

## 2. Superconducting single photon detectors in our DPS-QKD system

Figure 1(b) and (c) show the conceptual view and equivalent circuit of our SSPD system. We have developed a multi-channel NbN SSPD system[10, 11] based on the Gifford McMahon (GM) cryocooler that can offer guaranteed performance, be cryogen free, and be capable of turnkey, continuous operation with low input power consumption. There is a photograph of the SSPD system in the inset of on the left of Fig. 1(b). A superconductor nanowire has large inductance named kinetic inductance ($L_k$) in itself. We adopt a 50 Ω shunt resistance to suppress oscillation in the circuit shown in Fig. 1(b). The SSPD is biased with a current $I_b$ slightly below the critical value $I_c$. An incident photon generates a resistive "hotspot" which disrupts the superconductivity across the wire, resulting in propagating a voltage pulse to the AC coupled amplifier. The output voltage pulse of ~Gain×$I_b$× $R_L$ is input to the discriminator, where Gain is the gain of the amplifier. When a pulse height exceeds the threshold voltage of discriminator, the system registers the detection of photons

Potential time constants which would limit recovery time of a SSPD are (1) thermal diffusion[18], (2) time constant in the electrical circuit[19], and (3) latching[20]. In these time constants, (2) limits our SSPD system dominantly. In our system, the thermal conductance in the SSPD is high enough compared to input optical power. In addition, our SSPDs with active area of 15 μm×15 μm have $L_k$ with 1-2 μH and such a large $L_k$ decreases the probability of latching[20]. Thanks to such a large size, high coupling efficiency between a single mode fiber and an SSPD can be obtained. After the hotspot is formed, the superconducting current through it recovers with the time constant of $L_k/R_L$, where $R_L$ is the load resistance (in our case: 25 Ω). Since our SSPDs are susceptible to electrical reflection due to the impedance mismatch, we use a shunt resistance with 50Ω shown in Fig. 1(b) to avoid electrical oscillation or reflection. In the counter measure against a CW illumination attack, monitoring photocurrent is very useful and demonstrated in Ref. [21]. However, such a technique is not suitable for our system due to susceptibility to electrical oscillation. The current $I_b$ which flows in the superconductor nanowire depends on the time as follows.

$$I_b = I_0\left(1 - \exp(-R_L/L_k \, t)\right) = I_0\left(1 - \exp(-\beta t)\right), \quad (1)$$

where $I_0$ is the initial bias current. The detection efficiency (η) depends on the bias current. Ref [19] also investigated the time-dependence of the detection efficiency after the detection event by using optical pulse pair. Our goal in the future is to establish full quantum optical model of SSPDs, i.e., to obtain a full description of SSPD for positive operator valued measure (POVM) that can express a time dependent behavior for multiple photons. In the present work, we focus, as the first step, on studying the time dependent photo-detection probability for a pulse train. The standard model for the SSPD's POVM is given as,

$$\pi_{off} = \sum_{n=0}^{\infty}(1-\eta)^{n}|n\rangle\langle n|, \qquad (2)$$

where $\pi_{off}$ is the operator for no-detection event, $|n\rangle$ is for an n-photon state, and the detection operator is given $\pi_{on} = 1-\pi_{off}$. In this equation, we must introduce the function $\eta = \eta_{0}*f(I_b)$, where $\eta_0$ is the initial quantum efficiency. And $f(I_b)$ which describes a relative quantum efficiency dependent on bias current can be obtained from experimental results illustrated by example in Fig. 2. When we calculate $f(I_b)$, some fitting curves are assigned according to $I_b$. Decrease of the practical bias current according to eq. (1) induces the depression of the detection efficiency. Moreover, we have to consider the specific condition of the readout circuit of the SSPD.

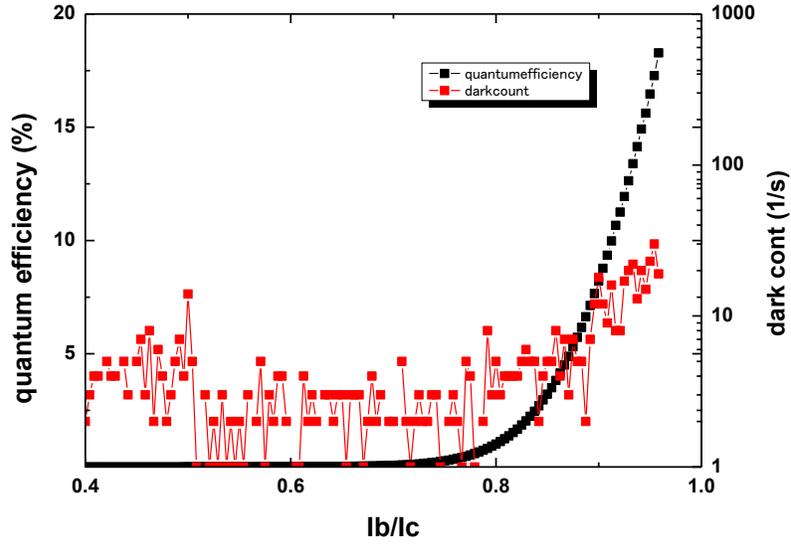

Fig. 2 Quantum efficiency and dark count rate of the superconductor single photon detector (CH5) as functions of bias current. Bias current ($I_b$) is normalized by critical current ($I_c$).

## 3. The detection probability of SSPDs for a pulse train

In this section, the detection probability for the train with an interval of T is considered. The input signal is set in coherent state $|\alpha\rangle$. The detection probability equation of an SPD for single event is described as "$1-\exp(-|\alpha|^2\eta)$" in Ref. [22]. To obtain the detection probability for the pulse train considering the recovery time of an SSPD, we decompose the detection process: At first, the normal conductance transition probability $S_{on}$ of SSPD to pulse train is obtained as follows in a recursive way.

$$S_{off}(1T) = e^{-\gamma}e^{-|\alpha|^2\eta_0}$$

$$S_{on}(1T) = 1 - S_{off}(1T)$$

$$S_{off}(2T) = e^{-\gamma}\left\{e^{-\gamma}e^{-|\alpha|^2\eta_0}e^{-|\alpha|^2\eta_0} + \left(1-S_{off}(1T)\right)e^{-|\alpha|^2\eta_0 f(I_0(1-\exp(-\beta T)))}\right\}$$

$$S_{on}(2T) = 1 - S_{off}(2T)$$

$$S_{off}(3T) = e^{-\gamma} \begin{Bmatrix} e^{-\gamma}e^{-2|\alpha|^2\eta_0}e^{-|\alpha|^2\eta_0} + \\ \left(1-S_{off}(1T)\right)e^{-\gamma-|\alpha|^2\eta_0 f(I_0(1-\exp(-\beta T)))}e^{-|\alpha|^2\eta_0 f(I_0(1-\exp(-2\beta T)))} \\ +\left(1-S_{off}(2T)\right)e^{-|\alpha|^2\eta_0 f(I_0(1-\exp(-\beta T)))} \end{Bmatrix}$$

$$S_{on}(3T) = 1 - S_{off}(3T)$$

..........  (3)

$$S_{off}(NT) = e^{-N\gamma}e^{-N|\alpha|^2\eta_0}$$
$$+\sum_{m=1}^{N-1}\left\{\left(1-S_{off}((N-m)T)\right)\left[\prod_{k=1}^{m}e^{-\gamma-|\alpha|^2\eta_0 f(I_0(1-\exp(-k\beta T)))}\right]\right\}$$

$$S_{on}(NT) = 1 - S_{off}(NT),$$

where $\gamma$ is the dark count probability of the SSPD, $S_{off}(NT)$ and $S_{on}(NT)$ are the superconducting state probability and normal conductance state probability at a time slot of "NT", respectively.

To recognize the photo-detection event in the SSPD system, $R_L \times I_b(NT) \times Gain$ must exceed threshold (Vth) of the discriminator. In our case the load resistance $R_L$ is 25Ω. Therefore, we have

$$P_{on}(NT) = S_{on}(NT)g(NT)$$
$$g(NT) = probability\_of\_(R_L \times I_b(NT) \times Gain \geq Vth) \quad (4)$$
$$P_{off}(NT) = 1 - P_{on}(NT),$$

where $P_{on}(NT)$ is the photo-detection event probability at the time slot of "NT". In our case, Vth is set around 8-40 m V. For example, Vth of CH5 is set 40 mV, and if $I_0$ (initial bias current: 22.2 μA) is added to the detector, 55.5 mV pulse is generated at photon detection. The minimum $I_b$ that can exceed Vth is about 72% of initial bias current ($I_0$). At that bias current, the detection efficiency η is 0.122% shown in Fig. 2. Note that we approximate g(NT) by $S_{off}((N-1)T)$ since this probability mostly affects the recovery of $I_b$.

### 4. Blinding SSPDs with pulse shaped illumination considering response time

Using these equations, behavior of the SSPD at blinding attack can be estimated. η, $L_k$ and $I_c$ of our SSPDs are listed in TABLE I. Figure 3 shows the counting rates as functions of input power of optical-pulse train of 300 ps width and 1 GHz repetition rate. Solid lines are experimental data and dashed lines are simulated values from eq. (4). In the simulated curves, a cutoff rate of the discriminator in our setting was taken into account. In the figure, one can see the plateau in the counting rate about 66-70 M count/s, which is the maximum counting rate of the discriminator in our setup. And the simulated figures are in agreement with the experimental values, especially in the high power incident region. Note that it is important for the eavesdropper to know the input power needed to blind SSPDs with count rate below a few hundred, since this is approximately the same as dark count rate of SSPDs connected to the field installed fibers so that Alice and Bob would regard these counts as the dark counts. Figure 3 indicates that the SSPD with a relatively small $L_k$ (CH5) requires input pulses of -30dBm to blind the SSPD with counting rate of a few hundred c/s. That means 8000 photons/pulse are necessary for blinding. And variation in $L_k$ makes more than 3dB difference in the optical-power for the blinding.

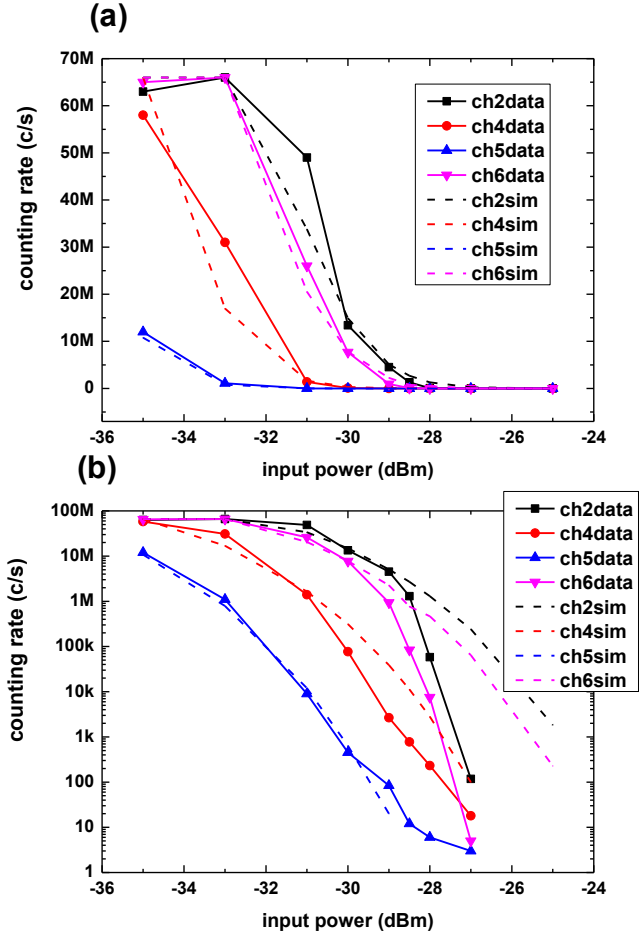

Fig. 3. Counting rate of SSPDs to pulse train of 1 GHz as functions of input power. Solid lines are experimental data and dashed lines are theoretical values from eq.(4). Linear (a), and log (b) scale.

Table 1. Characteristics of SSPDs.

| No.SSPD | η at dark count rate of 100c/s (%) | Critical current $I_c$ (μA) | Kinetic inductance $L_k$ (μH) |
|---|---|---|---|
| CH2 | 11.7 | 12.2 | 2.13 |
| CH4 | 12.1 | 24 | 1.14 |
| CH5 | 18.0 | 24.5 | 1.12 |
| CH6 | 10.0 | 13.1 | 1.73 |

While Lydresen et al.[15] used 0-π/2 modulated CW laser, Honjo et al.[16] employed 0-π modulated blinding pulse of 300 ps width at 1 GHz clock as shown in Fig. 4(a) and (b). The optical-input pulse for blinding decreased from mW to several μW. However, the required optical power for blinding can be reduced further, because the recovery time is much longer than the pulse interval of 1GHz repetition. As a result, many photons are required to blind SSPDs. To incapacitate SSPDs with weaker optical-power, we can deem the power of the optical pulse according to the calculation based on Eq. (4), as shown in Fig. 4(c). For example in CH5, the maximum η on which the $I_b$ is just below Vth is 0.122% mentioned in section 3. And the recovery time is calculated 58 ns by using eq. (1) and a kinetic inductance of CH5. Input pulse of 10000 photons per interval of 58 ns keeps CH5 at normal conductance state with more than 99.999% probability. In the case of CH2, the recovery time is is estimated as 13 ns and the required photons for blinding are 25000. To apply this consideration to the 1 GHz clock DPS-QKD system, which uses 1-nsec-double pulse, four times input power is necessary for blinding both of the detectors, assuming that two SSPDs have the same recovery time. Even in this case, the total optical-power for blinding is estimated only less than 90-990 nW, which is 5-10dB less than that of previous works. If a pair of SSPDs with different recovery times is used in a QKD system, 1-nsec-double pulse should be used to blind a target detector shown in Fig. 4(d). At the same time, bright light pulses are input to the other detector, and such a bright light pulse makes the other detector click with high probability. For example, suppose that CH2 and CH5 SSPDs are used in a DPS-QKD. When CH5 is blinded, CH2 clicks with probability of 89.4% for the first pulse and totally 98.9% at the first double pulse illumination. As a counter-measure to detect the blinding, Bob intends to have the simultaneous clicks of the detectors by switching the output port of the asymmetric Mach-Zehnder interferometer by changing its phase. In that case, the probability of NO-click in CH5 at the first double pulse and detection at the next pulse is estimated 0.22%, as a result, the probability of simultaneous detection in two ports increases to about 0.2% per blinding event. This value must be divided by the interval of 58 to compare with the normal 1GHz operating condition. As a result, the coincidence rate is estimated as 3.5e-3%, which is more than 29000 times larger than at normal operation. Therefore, Bob can detect the blinding pulse which is optimized for each SSPD by shifting the phase 0-π occasionally.

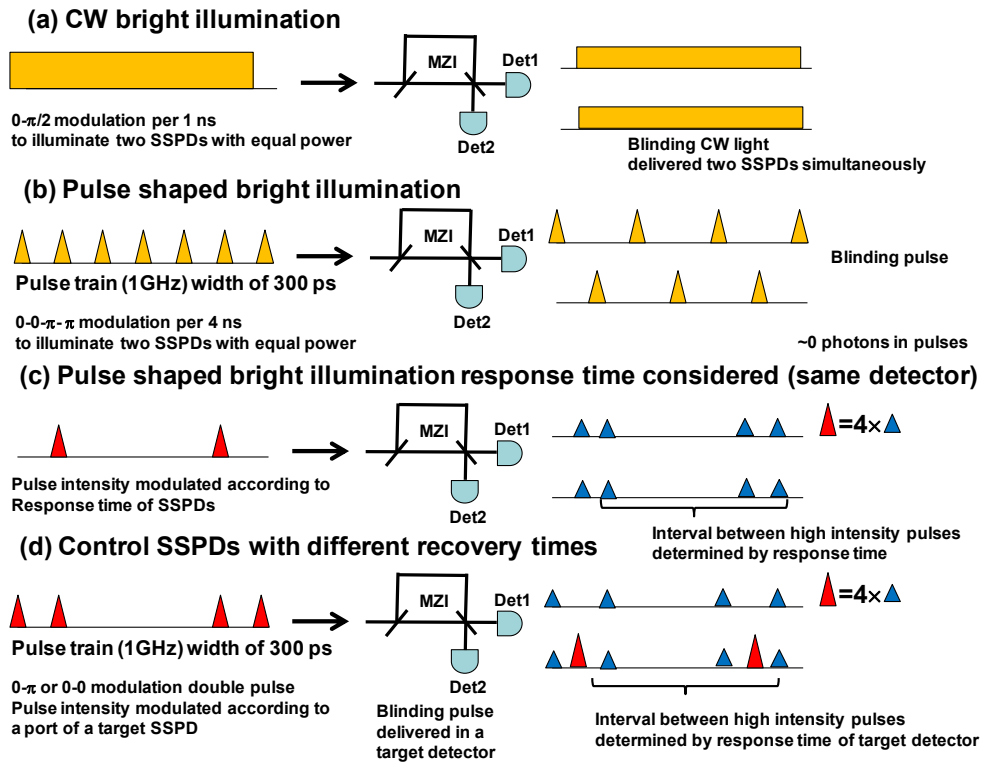

Fig. 4. Conceptual views of bright illumination blinding attack by (a) continuous wave laser, (b) pulse train, (c) pulse train with considering of response time, and (d) controlling detection port with 1-nsec-doble pulse.

## 5. Conclusion

In summary, we present an approach to obtain the equation which gives detection probability of SSPDs for a pulse train. A better understanding of the behavior under a pulse train, which our equation provides with us, is useful to consider an effective attack as well as countermeasure against bright illumination blinding attack to the QKD system. Such an analysis is necessary to dissect the security of the QKD system.